\documentclass[a4paper]{article}
\usepackage[dvips]{epsfig}
\usepackage{amssymb}
\usepackage{bm}
\usepackage{dcolumn}
\usepackage{amsmath}

\catcode`\@=11
\def\marginnote#1{}
\newcount\hour
\newcount\minute
\newtoks\amorpm
\hour=\time\divide\hour by60
\minute=\time{\multiply\hour by60 \global\advance\minute
by-\hour}\edef\standardtime{{\ifnum\hour<12
\global\amorpm={am}%
        \else\global\amorpm={pm}\advance\hour by-12 \fi
        \ifnum\hour=0 \hour=12 \fi
        \number\hour:\ifnum\minute<10
0\fi\number\minute\the\amorpm}}
\edef\militarytime{\number\hour:\ifnum\minute<10
0\fi\number\minute}

\def\draftlabel#1{{\@bsphack\if@filesw {\let\thepage\relax
   \xdef\@gtempa{\write\@auxout{\string
      \newlabel{#1}{{\@currentlabel}{\thepage}}}}}\@gtempa
   \if@nobreak \ifvmode\nobreak\fi\fi\fi\@esphack}
        \gdef\@eqnlabel{#1}}
\def\@eqnlabel{}
\def\@vacuum{}
\def\draftmarginnote#1{\marginpar{\raggedright\scriptsize\tt#1}}
\def\draft{\oddsidemargin -.5truein
        \def\@oddfoot{\sl preliminary draft \hfil
        \rm\thepage\hfil\sl\today\quad\militarytime}
        \let\@evenfoot\@oddfoot \overfullrule 3pt
        \let\label=\draftlabel
        \let\marginnote=\draftmarginnote

\def\@eqnnum{(\theequation)\rlap{\kern\marginparsep\tt\@eqnlabel}%
\global\let\@eqnlabel\@vacuum}  }

\def\numberbysection{\@addtoreset{equation}{section}
        \def\theequation{\thesection.\arabic{equation}}}

\def\underline#1{\relax\ifmmode\@@underline#1\else
 $\@@underline{\hbox{#1}}$\relax\fi}

\catcode`@=12
\relax

\numberbysection

\advance \topmargin by -\headheight
\advance \topmargin by -\headsep

\evensidemargin \oddsidemargin
\def\br{\begin{eqnarray}}
\def\er{\end{eqnarray}}
\def\be{\begin{equation}}
\def\ee{\end{equation}}

\def\({\left(}
\def\){\right)}

\relax

\def\D{\Delta}

\def\pa{\partial}

\def\tp0{\Theta_{+}^{(0)}}
\def\tm0{\Theta_{-}^{(0)}}

\def\f#1#2#3 {f^{#1#2}_{#3}}

\def\win1{{\sf w_{1+\infty}}}

\def\Win1{{\sf W_{1+\infty}}}

\def\rlx{\relax\leavevmode}
\def\inbar{\vrule height1.5ex width.4pt depth0pt}
\def\IZ{\rlx\hbox{\sf Z\kern-.4em Z}}
\def\IR{\rlx\hbox{\rm I\kern-.18em R}}
\def\IC{\rlx\hbox{\,$\inbar\kern-.3em{\rm C}$}}
\def\IN{\rlx\hbox{\rm I\kern-.18em N}}
\def\IO{\rlx\hbox{\,$\inbar\kern-.3em{\rm O}$}}
\def\IP{\rlx\hbox{\rm I\kern-.18em P}}
\def\IQ{\rlx\hbox{\,$\inbar\kern-.3em{\rm Q}$}}
\def\IF{\rlx\hbox{\rm I\kern-.18em F}}
\def\IG{\rlx\hbox{\,$\inbar\kern-.3em{\rm G}$}}
\def\IH{\rlx\hbox{\rm I\kern-.18em H}}
\def\II{\rlx\hbox{\rm I\kern-.18em I}}
\def\IK{\rlx\hbox{\rm I\kern-.18em K}}
\def\IL{\rlx\hbox{\rm I\kern-.18em L}}
\def\one{\hbox{{1}\kern-.25em\hbox{l}}}
\def\0#1{\relax\ifmmode\mathaccent"7017{#1}%
B        \else\accent23#1\relax\fi}

\def\PRL#1#2#3{{\sl Phys. Rev. Lett.} {\bf#1} (#2) #3}
\def\NPB#1#2#3{{\sl Nucl. Phys.} {\bf B#1} (#2) #3}

\def\JMP#1#2#3{{\sl J. Math. Phys.} {\bf #1} (#2) #3}

\def\JHEP#1#2#3{{\sl JHEP} {\bf #1} (#2) #3}

\def\IJMPB#1#2#3{{\sl Int. J. Mod. Phys. B} {\bf #1} (#2) #3}
\def\PHYSD#1#2#3{{\sl Physica D: Nonlinear Phenomena} {\bf #1} (#2) #3}
\def\NATCOMM#1#2#3{{\sl nature communications} {\bf #1} (#2) #3}  
\def\SCIREP#1#2#3{{\sl Scientific Reports} {\bf #1} (#2) #3}  
\def\MATH#1#2#3{{\sl Mathematics} {\bf #1} (#2) #3}  

\begin{document}
\begin{titlepage}

\vskip .6in

\begin{center}
{\large {\bf Modified open Toda chain and quasi-integrability}}
\end{center}

\normalsize
\vskip .4in

\begin{center}

H. Blas, A. C. R. do Bonfim, L. T. Teixeira, and G.K.R. de Souza 

\par \vskip .2in \noindent

Instituto de F\'{\i}sica\\
Universidade Federal de Mato Grosso\\
Av. Fernando Correa, $N^{0}$ \, 2367\\
Bairro Boa Esperan\c ca, Cep 78060-900, Cuiab\'a - MT - Brazil. \\ 
\normalsize
\[\]
Corresponding author: Harold Blas
\end{center}

\par \vskip .3in \noindent
We present a study of a quasi-integrable deformation of the three-particle open Toda chain, constructed by introducing a translation-invariant three-body interaction terms. Although this modification explicitly breaks the exact integrability of the standard Toda model, it retains fundamental structural properties, including energy and momentum conservation. Furthermore, we show that under a specific time-reflection and discrete symmetry among the chain coordinates, the system admits a quasi-conserved higher-order integral. Through analytic and numerical analysis of the deformed dynamics, we demonstrate the emergence and long-time persistence of quasi-conserved quantities, thereby establishing a controlled realization of quasi-integrability in a minimal nonlinear chain. Given the central role of integrable systems in elucidating the dynamics of classical and quantum models, this framework provides a concrete setting to investigate the mechanisms underlying the gradual breakdown of integrability and the onset of quasi-integrability in few-body systems.
\end{titlepage}

\section{Introduction}

The Toda chain has long been recognized as a paradigmatic model in the theory of integrable systems \cite{toda1}. Its exponential nearest-neighbor interactions lead to an exactly solvable dynamics, admitting a Lax pair representation, an infinite hierarchy of conserved quantities, and soliton solutions with remarkable stability properties. The Toda chain serves as a unifying framework for understanding nonlinear collective behavior, energy localization, and integrability across a wide range of physical systems-from crystal lattices to quantum field theories. For the case of three particles, the Toda chain provides a minimal nontrivial setting where integrability manifests in a transparent manner: the system supports three independent integrals of motion in involution, usually taken to be the total momentum, the Hamiltonian, and a third higher-order invariant. 

Despite its celebrated integrability, the Toda chain in its standard form only accounts for nearest-neighbor interactions. In many physical contexts, however, nonlinear chains are subject to corrections beyond this idealized structure. These include next-to-nearest neighbor couplings, multi-particle correlations, or effective forces emerging from higher-order interactions. Introducing such modifications allows one to probe the robustness of integrability, to construct quasi-integrable deformations, and to explore routes to chaos in systems that are otherwise exactly solvable. 

In this work we consider a specific deformation of the three-particle Toda chain, obtained by supplementing the model with a three-body interaction potential. We will choose the additional potential to preserve translation invariance, and thus the conservation of total momentum, while it generically breaks the integrable structure associated with the Toda hierarchy. To the best of our knowledge, the deformations of the Toda chain have mainly considered two-particle deforming potentials (see e.g. \cite{bilman}). For an integrable deformation  case of three-particle system with two-body interactions governed by Toda forces but modified by the addition of new self-interaction terms, see \cite{ranada}.   

Higher-order interactions constitute a fundamental aspect of numerous complex systems. In ecological networks, for example, the interaction between two species may be modulated by the presence of a third; in social systems, collective phenomena naturally arise from interactions involving groups of three or more individuals; and in neural networks, such as those in the cerebral cortex, non-pairwise couplings play a critical role in information processing. Recent advances employing mathematical frameworks such as simplicial complexes and hypergraphs have demonstrated that the dynamical behavior induced by higher-order interactions can differ markedly from that of systems governed exclusively by pairwise couplings. Consequently, the identification and modeling of higher-order effects have become central to understanding the emergent dynamics and functional organization of complex many-body systems. In Ref. \cite{malizia}, the authors proposed a method for the reconstruction of the structural connectivity in systems composed of coupled non-linear oscillators, enabling the identification of both pairwise and higher-order interactions directly from the system’s temporal evolution.
 
On the other hand, quasi-integrable models represent a class of deformations applied to exactly integrable field theories, such as sine-Gordon, KdV, and Toda. While not fully integrable, these models preserve several essential characteristics, including a hierarchy of (asymptotically) conserved quantities and soliton-like solutions. The theoretical foundation for quasi-integrability was first established via an anomalous Lax formulation \cite{fer1, fer2}. It is achieved by modifying the potential of the temporal Lax component or the corresponding Hamiltonian \cite{kumar}. Subsequently, an alternative description based on a pseudo-potential approach was developed \cite{npb1, jhep, ijmpb2, math1}, providing another perspective on the subject.  

Within this broader perspective, the modified $N=3$ open Toda chain offers a minimal yet representative model for exploring how higher-order interactions influence integrable dynamics and the transition toward quasi-integrable or non-integrable regimes.

In many-body systems, the introduction of a perturbation generally breaks integrability, resulting in the preservation of only a limited number of conserved quantities. So, our analysis first establishes the exact conserved quantities that survive in the presence of the three-body deformation, namely the Hamiltonian (energy) and the total momentum. We then outline possible strategies to preserve the higher Toda integral of motion in the regime of three-body coupling. This approach offers a systematic and controlled framework for investigating the deformation of integrable structures and the subsequent emergence of quasi-conserved quantities, as observed in deformed soliton models \cite{math1, jhep}.

Earlier studies of lattice solitary waves and their collisions (see, e.g., \cite{hietarinta}) demonstrated that, in a dissipative Toda lattice, solitary waves develop decay and tails, yet head-on collisions of two solitons with different amplitudes remain effectively elastic. This indicates that key soliton characteristics may persist in nonconservative lattice systems, anticipating quasi-integrable behavior. Extending deformation analyses of integrable models, we develop within the Toda-lattice framework a previously unexplored approach demonstrating that certain non-integrable theories can preserve key structural features of fully integrable systems. Specifically, we formulate and analyze quasi-integrability in a minimal three-particle Toda-type chain with a genuine three-body deformation and an explicit anomaly-cancellation mechanism.

The paper is organized as follows. The next section defines the modified Toda chain. It also discusses the equations of motion, the exact and quasi-conservation laws. In section \ref{sec:quasi} we discuss the quasi-conserved charge and vanishing anomaly. In section \ref{sec:hirota} we present an exact solution of the Toda chain, satisfying the time-reflection symmetry. The section \ref{sec:expansion}
presents an expansion around the $N=3$ Toda chain for a specific deformed potential. The vanishing of the anomaly is verified. In section \ref{sec:num} we support our results by means of numerical simulation of the modified open Toda model. Sec. \ref{sec:conclu} presents our conclusions and discussions. The appendix \ref{sec:apprelax} presents the algorithm suitable for solving the model 
using the relaxation method. 

\section{A modified $N=3$ open Toda model}

The classical Toda chain describes a one-dimensional chain of particles interacting through an exponential nearest-neighbor potential \cite{toda1}. 
In the present work we consider the minimal nontrivial case  $N=3$, and introduce some deformations that incorporate a three-body interaction. The modified Hamiltonian reads
\br
\label{3body}
H= \sum_{i=1}^3 \frac{p_i^2}{2} + e^{q_1-q_2} + e^{q_2-q_3} + U_3(q_1, q_2, q_3),
\er
where the new term $U_3(q_1, q_2, q_3)$ is given by
\br
\label{poterms}
U_3(q_1, q_2, q_3) &=& U^{(1)}(y_1) +U^{(2)}(y_2) +U^{(3)}(y_3),\\
y_1&=&\mu_1 q_{2}-(\mu_1+\nu_1) q_1+ \nu_ 1 q_{3},\\
 y_2&=& \mu_2 q_{1}-(\mu_2 +\nu_2) q_2 + \nu_2 q_{3},\\
 y_3&=& \mu_3 q_{1}-(\mu_3+\nu_3) q_3+ \nu_3 q_{2},\,\,\,\,\,\,\, \mu_i,\nu_i \in \IR.
\er
here, $U^{(i)}(y_i)$ represent a generic translation invariant potential. The arguments $y_i$ correspond to weighted discrete `second derivative' quantities  of the particle positions, i.e. the lattice “curvature” around each $q_i$.
  
This deformation is natural for several reasons. First, it respects translational invariance of the system, since the interaction depends only on the coordinate differences $y_i$; as a consequence, total momentum remains an exact conserved quantity. Second, it introduces a cooperative interaction among all three particles that cannot be reduced to a superposition of pairwise forces, thereby going beyond the conventional Toda chain structure.

In the limit $U_3=0$ the Hamiltonian reduces to the standard three-particle Toda model, which is integrable and admits three independent constants of motion in involution. For finite $U_3$, however, the presence of the three-body terms generically breaks the Toda integrability, leaving only the energy and momentum conservation laws guaranteed by symmetry. This setup thus provides a minimal laboratory for investigating how multi-particle interactions deform the integrable structure and give rise to quasi-integrability, under certain conditions.
 
The dynamics of the three-particle system follow from Hamilton’s equations,
\br
\dot{q}_i = \frac{\pa H}{\pa p_i},\,\,\,\,\dot{p}_i = -\frac{\pa H}{\pa q_i},\,\,\,\, i=1,2,3.
\er
For the Hamiltonian defined in Eq. (\ref{3body}), the coordinate equations of motion are simply
\br
\label{dq1}
\dot{q}_i = p_i,\,\,\,\,\, i=1,2,3.
\er
The momentum equations become
\br 
\dot{p}_1 &=& - V'_T(q_1-q_2) + (\mu_1+\nu_1) U^{(1)'}(y_1) - \mu_2 U^{(2)'}(y_2) - \mu_3 U^{(3)'}(y_3),\label{dp1}\\
\label{dp2}
\dot{p}_2 &=& V'_T(q_1-q_2) - V'_T(q_2-q_3) - \mu_1 U^{(1)'}(y_1) +(\mu_2+\nu_2) U^{(2)'}(y_2) - \nu_3 U^{(3)'}(y_3),\\
\label{dp3}
\dot{p}_3 &=&  V'_T(q_2-q_3) - \nu_1 U^{(1)'}(y_1) - \nu_2 U^{(2)'}(y_2) + (\mu_3+\nu_3) U^{(3)'}(y_3),
\er 
with the following definitions 
\br
V'_T(y) &\equiv & \frac{d}{dy} V_T(y),\,\,\,\,\ V_T(y) = e^{y},\\
U^{(i)'}(y_i) &\equiv & \frac{d}{dy_i} U^{(i)}(y_i),\,\,\,\, i=1,2,3.   
\er 
The equations (\ref{dp1})-(\ref{dp3}) show the interplay between the conventional Toda nearest-neighbor forces and the additional collective force generated by the three-body coupling. Note that the force $U^{(i)'}(y_i)$ enters weighted by ($\mu_i + \nu_i $) for the $i'th$ particle and with opposite signs $-\mu_j$ and $-\nu_j$ ($j \neq i$) for the other particles.
 
Let us define the center of mass (CM) coordinate as 
\br
\label{CM1}
Q_{CM} \equiv \frac{1}{3} (q_1+q_2+q_3).
\er
So, one can readily confirm that the internal forces add up to zero, i.e. using $p_i=\frac{d q_i}{dt}$ and the system of equations (\ref{dp1})-(\ref{dp3}) one has 
\br
\label{CM2}
\frac{d^2}{dt^2} Q_{CM} =0\,\,\,\, \rightarrow\,\,\,\, Q_{CM} = v_o t + Q_o.
\er
Therefore, the net force on the center of mass is zero, which guarantees momentum conservation and confirms the physical consistency of the three-body interaction.

\subsection{Conserved quantities: usual Toda chain}
\label{sec:cons}

The integrable structure of the Toda chain is characterized by the existence of a complete set of independent conserved quantities in involution. For a chain of three particles, this includes the total momentum, the Hamiltonian, and a third nontrivial invariant of cubic order in the momenta. The introduction of the exponential three-body interactions, however, alters this picture in a fundamental way.

The Hamiltonian of Eq. (\ref{3body}) is translationally invariant: it depends only on coordinate differences $(q_{i+1}-q_i)$ and on the lattice curvatures $y_i$, both of which remain unchanged under uniform shifts 
\br
q_i \rightarrow q_i + c.
\er
By Noether’s theorem, this symmetry ensures the exact conservation of total momentum,
\br
\label{mom}
\dot{P}=0, \,\,\,\,\,\,\,P \equiv  p_1+p_2+p_3.
\er
Furthermore, since the dynamics is generated by a time-independent Hamiltonian, the total energy is preserved,
\br
\label{ener}
\dot{H} =0.
\er

For the usual Toda chain one has a third conserved charge
\br
\label{I3}
\dot{I}_3 =0,\,\,\,\,\,\,\,\,\,I_3 \equiv p_1^3 + p_2^3 +p_3^3 + 3 [ p_1 e^{q_1-q_2} + p_2 (e^{q_1-q_2} +e^{q_2-q_3} ) + p_3 e^{q_2-q_3}].
\er 

Below, we examine the behavior of the charge of type $I_3$ under the modified model.
  
\subsection{Exact quasi-conservation law}

Next we examine an exact expression for the time derivative of the Toda cubic invariant ($I_3$) when the three-body interaction is present, and identify the anomaly explicitly.

Taking the time-derivative of $I_3$ in (\ref{I3}) and using the equations of motion (\ref{dp1})-(\ref{dp3}), one can write
\br
\label{ano1}
\frac{1}{3} \dot{I}_3 &\equiv& {\cal A}_3(t)\\
\nonumber
&=& [(\mu_1+\nu_1) p_1^2 -\mu_1 
p_2^2 -\nu_1 p_3^2 +\nu_1 e^{q_1-q_2} -(\mu_1+\nu_1) e^{q_2-q_3}) ] \frac{d}{dy_1} U^{(1)}(y_1) +\\
&&[-\mu_2 p_1^2 + (\mu_2+\nu_2) 
p_2^2- \nu_2 p_3^2 + \nu_2 e^{q_1-q_2} + \mu_2  e^{q_2-q_3} ] \frac{d}{dy_2}U^{(2)}(y_2)+\label{quasi1} \\
&&[-\mu_3 p_1^2 -\nu_3 
p_2^2  + (\mu_3+\nu_3) p_3^2 -(\mu_3+\nu_3) e^{q_1-q_2} + \mu_3 e^{q_2-q_3} ] \frac{d}{dy_3} U^{(3)}(y_3).\nonumber
\er
Therefore, the cubic Toda invariant $I_3$ satisfies the exact non-  perturbative quasi-conservation law (\ref{ano1})-(\ref{quasi1}) for the modified model (\ref{3body}). This identity is algebraic and exact: no series expansion, no small-$U^{(i)}$ assumption, and no truncation have been used. The quantity ${\cal A}(t)$ defined in (\ref{ano1}), referred as the anomaly, quantifies the degree of deformation of the charge $I_3$ induced by the potential $U_3$.  

\section{Quasi-conserved charge and vanishing anomaly}
\label{sec:quasi}

A quasi-integrable model, as introduced in \cite{fer1, npb1, jhep} (see also the review paper \cite{math1}) designates a class of deformations of exactly integrable field theories-such as the sine-Gordon, Korteweg–de Vries (KdV), or Toda models-that retain several fundamental characteristics of integrability, including the presence of a hierarchy of (asymptotically) conserved quantities and soliton-like configurations. The anomalous Lax formulation of  quasi-integrability  has been presented in \cite{fer1, fer2}, whereas the pseudo-potential approach in \cite{npb1, jhep, ijmpb2}.

In the framework of quasi-integrable field theories, the space–time reflection symmetry plays a crucial role in ensuring the asymptotic conservation of charges and the emergence of quasi-integrable behavior.  
More precisely, although the deformed theory no longer satisfies the exact zero-curvature condition of the integrable case, it admits a deformed Lax representation in which a curvature anomaly term 
$\chi(x,t)$ appears. This anomaly governs the violation of integrability and directly affects the time evolution of the conserved charges
\br
\frac{d \widetilde{Q}_n}{dt}  = \int_{-\infty}^{+\infty} \, dx\, \chi_n,\,\,\,\, n=1,2,3,...
\er
The space–time reflection symmetry,
\br
(x,t) \rightarrow -(x,t),
\er
ensures that the anomaly density $\chi_n$ behaves as an odd function under this transformation, while the integrand or measure remains even. Consequently, the space-time integral of the anomaly vanishes, leading to
\br
\D \widetilde{Q}_n = \widetilde{Q}_n(+\infty)-\widetilde{Q}_n(-\infty) =0,
\er
even though $\dot{\widetilde{Q}}_n \neq 0$ locally.

Thus, the reflection symmetry acts as a compensating mechanism: it guarantees that the total anomaly cancels between symmetric regions of space–time, yielding asymptotic charge conservation and elastic soliton scattering. In this sense, the preservation (or restoration) of space–time parity symmetry is a key structural ingredient that allows a deformed, non-integrable model to exhibit quasi-integrable dynamics, bridging the gap between exact integrability and generic nonlinear behavior.

In the present setting, the model is a purely mechanical system with a finite number of degrees of freedom. The absence of a continuous spatial variable implies that the dynamics is governed by a discrete set of particle coordinates rather than by field configurations. Within this framework, quasi-integrability will manifest itself through asymptotic in time conservation laws and residual symmetry properties of the few-body dynamics, thereby providing a finite-dimensional mechanical analogue of the quasi-integrability concepts originally developed in field-theoretical contexts.

\subsection{Quasi-integrability in deformed $3-$particle open Toda lattice}

Consider the Hamiltonian (\ref{3body}), viewed as a deformation of the integrable open three-particle Toda lattice. Let $I_{j}(t),\, j = 1,2,3$, denote the Hamiltonian ($I_1 = H$) and the total momentum ($H_2 = P$) conserved quantities of the deformed Toda model, whereas the third $I_3$ quantity is defined in the same functional form as in (\ref{I3}), although it now refers to the deformed (non-integrable) Toda lattice.

In the deformed system, these quantities generally satisfy evolution equations of the form
\br
\frac{d}{dt} I_{j}(t) = {\cal A}_{j}(t),\,\,\,\, j=1,2,3,
\er
where ${\cal A}_{j}(t)$ are known as anomaly functions induced by the deformation. In the case at hand, one has
\br
{\cal A}_{j}(t) =0, \,\,\,\, j=1,2,
\er 
which correspond to the vanishing of the anomalies for the exact conservation laws (\ref{mom}) and (\ref{ener}) of the energy (${\cal A}_1=0$) and momentum (${\cal A}_2=0$), respectively. 

The system is said to be quasi-integrable if there exists a nontrivial class of trajectories for which the third integrated anomaly vanishes asymptotically,
\br
\label{anoint1}
\int_{-t_0}^{t_0} \,dt\, {\cal A}_{3}(t) = 0\,\, \mbox{as}\,\,  t_0 \rightarrow +\infty,
\er
so that the corresponding charge $I_{3}(t)$ remain the same in the past and in the future, despite not being conserved locally in time. The central mechanism responsible for the vanishing of the integrated anomaly, as we will see below, is a discrete symmetry combining time reversal and particle relabeling. 

Let us summarize the notion of quasi-integrability adopted in the above $N=3$ Toda lattice setting.

1) If we have a set of solutions of the system of equations (\ref{dp1})-(\ref{dp3}) transforming under the time-reflection and coordinate relabeling  ${\cal T}_{cr}$ as 
\br
\label{tcr}
{\cal T}_{cr} :  t \rightarrow -t,\,\,\,\, q_1(t) \leftrightarrow - q_3(-t),\,\,\,\,\,q_2(t) \rightarrow  - q_2(-t),  
\er
with the corresponding transformation of momenta and auxiliary variables.

2) And if the potential term $U_3(q_1,q_2,q_3)$, evaluated on such a solution is even under the parity ${\cal T}_{cr}$, i.e.
\br
\label{U3s}
{\cal T}_{cr} (U_3) = U_3,
\er
such that (\ref{anoint1}) holds true,

3) Then we have the quantity $I_3(t)$ to be conserved asymptotically, i.e. 
\br
 I_3(t_0) = I_3(-t_0).
\er
Theories possessing such properties we call quasi-integrable $N=3$ open Toda lattice theories.

Under the transformations (\ref{tcr})-(\ref{U3s}) the equations of motion are invariant, while the anomaly function ${\cal A}_{3}(t)$ associated with the broken conserved charge transforms as
\br
\label{odd1}
{\cal A}_{3}(-t)= - {\cal A}_{3}(t),
\er
i.e. it possesses odd parity. As a consequence, the integrated anomaly over any symmetric time interval vanishes identically (\ref{anoint1}).
This odd-parity property is therefore sufficient to guarantee asymptotic conservation of the corresponding charge. These symmetries are sufficient, but not generic, and their presence delineates the quasi-integrable sector of the deformed Toda dynamics.

The vanishing  of the integrated anomaly relies on discrete parity symmetries that impose mirror-symmetric constraints on the initial or asymptotic lattice configurations, thereby selecting parity-invariant sectors of phase space; for generic asymmetric configurations these constraints are not satisfied, the anomaly lacks definite parity, and the asymptotic conservation of the $I_3$ Toda charge is lost.

At the level of initial conditions, parity invariance requires that particle positions and momenta be arranged in a mirror-symmetric fashion with respect to the center of the chain, such that
\br
q_i(0) = - q_{4-i}(0),\,\,\,p_i(0) = p_{4-i}(0),\,\,\,\, i=1,2,3. 
\er
These relations define a lower-dimensional submanifold of the full phase space and are therefore not satisfied by generic initial data.

Next, following the above reasoning let us integrate (\ref{ano1}) in time from $t_i\equiv-t_0$ to $t_f \equiv +t_0$
\br
\label{ano011}
\frac{1}{3} \int_{-t_0}^{t_0} \, dt \,\dot{I}_3 &=&\int_{-t_0}^{t_0} \,dt\, {\cal A}_3(t),\\
\D I_3   
 &\equiv&  \frac{1}{3}(I_3(t_0)-I_3(-t_0)). \label{ano11}
\er
Let us impose the condition 
\br
\label{vanish}
\int_{-t_0}^{t_0} \,dt\, {\cal A}_3(t)=0.\er
Then, from (\ref{ano011})-(\ref{ano11}) and (\ref{vanish}) one obtains 
\br
\label{i30}
\D I_3 =0 \rightarrow  I_3(t_0) = I_3(-t_0).
\er 
In the limit $t_0 \rightarrow \infty$, one defines $I_3$ as an asymptotically conserved charge which characterizes the $N=3$ deformed Toda chain as a quasi-integrable model. This charge is conserved   asymptotically, even though $\dot{I}_3 \neq 0$ locally.  

By carefully examining the anomaly expressions (\ref{quasi1}) one concludes that the odd parity condition (\ref{odd1}) can be obtained for the lattice coordinates satisfying the symmetry
\br
\label{sym1}
q_3(-t) &=& -q_1(t),\,\,\,\,\,q_2(-t) = - q_2(t),\\
 y_i(-t) &=&- y_i(t),\label{sym2}
\er
and the deformation potential terms (\ref{poterms}) with even symmetry, i.e.
\br
\nonumber
U^{(i)}(y_i(-t)) &=& U^{(i)}(-y_i(t))\\
&=& U^{(i)}(y_i(t)),\label{potsymm1}
\er
provided that the parameters satisfy the relationships
\br
\label{param1}
\mu_1 = - 2 \nu_1,\,\,\,\,\mu_2 = \nu_2,\,\,\,\,\,\nu_3 = - 2 \mu_3.
\er
Remarkably, the symmetry (\ref{sym1})-(\ref{sym2}) associated with the deformed Toda chain can be regarded as the discrete counterpart of the continuous space-reflection symmetry $x \rightarrow -x$ that characterizes the quasi-integrable soliton models extensively investigated in the literature. In fact, the time and discrete coordinate reflection symmetry of the $N=3$ open Toda model can be written as
\br
(q_1, q_2, q_3;t) \rightarrow - (q_3, q_2, q_1;t).
\er
 
Quasi-integrability, together with the associated notion of asymptotically conserved charges, was introduced in Refs. \cite{fer1}-\cite{ijmpb2}. In this setting, quasi-conservation laws are not generic properties of the model but hold only for restricted classes of solutions-e.g., kink-antikink and breather configurations in deformed sine-Gordon theories, or multi-soliton states in deformed nonlinear Schr\"{o}dinger models with vanishing (bright) or nonvanishing (dark) boundary conditions. In all such cases, field configurations possess definite parities under space-time reflections. This contrasts with conventional integrability, where conserved charges are globally defined on phase space and preserved for arbitrary configurations. An analogous mechanism arises in the finite-dimensional Toda deformation considered here: a translation-invariant three-particle interaction breaks global Toda integrability, while a remnant of the original conserved quantities is only asymptotically conserved for specific initial data or scattering regimes. Thus, the model furnishes a finite-dimensional realization of quasi-integrability.

\section{Hirota-Moser solutions}
\label{sec:hirota}

Let us carefully analyze the symmetry (\ref{sym1}) for the integrable three-particle Toda chain solution. The usual $N=3$ Toda chain equations of motion become
\br
\label{t1}
\frac{d^2}{dt^2} q_1(t) &=& - e^{q_1(t)-q_2(t)},\\
\label{t2}
\frac{d^2 }{dt^2} q_2(t) &=&  e^{q_1(t)-q_2(t)} -  e^{q_2(t)-q_3(t)},\\
\frac{d^2}{dt^2} q_3(t)&=& e^{q_2(t)-q_3(t)}. \label{t3}
\er
The physically relevant relative exponentials (and hence the interparticle distances) are expressed in terms of the tau functions by the standard Hirota-Moser relations \cite{moser, toda2}
\br
e^{q_1(t)-q_2(t)} = \frac{\tau_0(t)\tau_2(t)}{\tau_1(t)^2},\,\,\,\,\,e^{q_2(t)-q_3(t)} = \frac{\tau_1(t)\tau_3(t)}{\tau_2(t)^2}.
\er  
Consider the tau functions
\br
\tau_1(t) = \tau_2(t)  = \frac{1}{2} (1+ \cosh{(\sqrt{2}\, t)}),\,\,\,\,\tau_0(t) = \tau_3(t) =1. 
\er
Then, a solution satisfying the symmetry (\ref{sym1}) becomes
\br
\label{sol1}
q_1(t) = - \log{\tau_1(t)},\,\,\,\,q_3(t) = \log{\tau_1(t)},\,\,\,\,
q_2(t) = 0.
\er 
One can directly verify that this solution satisfies the usual $N=3$ Toda chain equations (\ref{t1})-(\ref{t3}). The solution (\ref{sol1}) is plotted in the Fig.1 showing the symmetry (\ref{sym1}). 

\section{The expansion around the $N=3$ Toda chain}
\label{sec:expansion}

The construction of the quasi-conserved charge $I_3$ presented above was carried out for a broad class of potentials, without providing explicit estimates for the magnitude of the anomaly associated with $I_3$. We now address the evaluation of the anomaly, defined in Eq. (\ref{ano1})-(\ref{quasi1}), and assess the effectiveness of the quasi-conservation law (\ref{vanish})-(\ref{i30}). To this end, we consider a specific potential that represents a perturbation of the $N=3$ Toda potential while exhibiting the essential features of the deformation, such as the symmetry (\ref{potsymm1}) and describing the three-body interaction.

\begin{figure}
\centering
\label{fig1k}
\includegraphics[width=1.5cm,scale=4, angle=0,height=5cm]{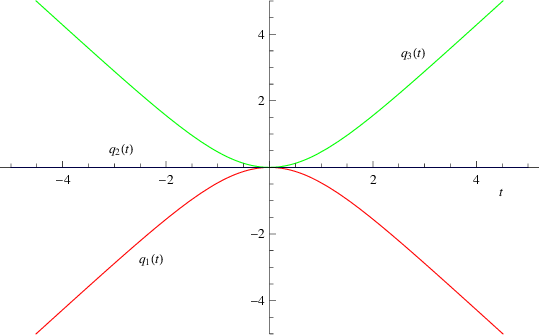} 
\parbox{6in}{\caption{(color online) A solution of the $N=3$ Toda chain coordinates $q_1(t)$ (red), $q_2(t)$ (blue) and $q_3(t)$ (green). Notice that the symmetry (\ref{sym1}) is realized.}}
\end{figure}
Taking into account (\ref{param1}) one can show that 
\br
y_1 = -\frac{1}{2} \mu_1 (q_1 - 2 q_2 + q_3),\,\,\,\,y_2 = \mu_2 (q_1 - 2 q_2 + q_3),\,\,\,\,
y_3 = \mu_3 (q_1 - 2 q_2 + q_3),\,\,\,\,
\er
Then, one considers a general deformation potential of type 
\br
\label{U3y}
U_3(y) &=& U^{(1)}(y_1) + U^{(2)}(y_2) + U^{(3)}(y_3),
\er
where
\br
\label{Q3y} 
Q_3 &\equiv& (q_1 - 2 q_2 + q_3),\,\,\,\, y_i = \zeta_i Q_3,\\
U^{(i)}(-y_i) &=& U^{(i)}(y_i),\,\,\,\,\,\,\,
\zeta_1= -\frac{\mu_1}{2},\,\,\zeta_2 = \mu_2,\,\, \zeta_3 = \mu_3.
\er
Let us choose the next symmetric potential as required by (\ref{U3s}) or (\ref{potsymm1})
\br
\label{Ui}
U^{(i)}(y) \equiv  \cosh{(\zeta_i Q_3)}-1.
\er
So, the quasi-conservation law (\ref{ano1})-(\ref{quasi1}) can be written as 
\br
\label{ano2}
\frac{1}{3} \dot{I}_3 &\equiv& {\cal A}_3(t)\\
 &=& - [p_1^2 -2 p_2^2 + p_3^2 - e^{Q_1} -e^{Q_2}] \Big[\zeta_1 \sinh{(\zeta_1 Q_3)} +\zeta_2 \sinh{(\zeta_2 Q_3)} +\zeta_3 \sinh{(\zeta_3 Q_3)} \Big], 
 \label{quasi2}\\
\label{Q12} 
Q_1 &\equiv& q_1 - q_2,\,\,\, Q_2 \equiv q_2 - q_3.
\er
Notice that the symmetry condition (\ref{sym1})-(\ref{sym2}) implies 
\br
p_1^2 &\rightarrow & p_3^2,\,\,\,\,\,\,p_2^2 \rightarrow p_2^2\\
Q_1(-t) &=& Q_2(t),\,\,\,\, Q_3(-t) = -Q_3(t).
\er
Then, one has an anomaly with  odd parity, i.e. ${\cal A}_3(-t)= -{\cal A}_3(t)$.

To the best of our knowledge, the interaction introduced in Eq. (\ref{Ui}) has not been considered in previous studies of the Toda lattice or its deformations. Classical Toda systems and their integrable extensions rely on nearest-neighbor exponential interactions associated with Lie-algebraic structures and related $q$-deformations. While integrable deformations with Toda-type two-body forces and additional self-interaction terms are known (see, e.g., \cite{ranada}), they remain within a two-body framework and do not incorporate the translation-invariant three-particle cosh-type interaction studied here. The present deformation therefore constitutes a distinct modification within the class of Toda-type systems.

\subsection{Dynamics versus parity}

We examine the conditions under which the aforementioned criteria can be fulfilled, thereby leading to quasi-integrable behavior. We restrict our analysis to the class of theories with potential deformation  terms of type (\ref{Ui}); that is, to models whose equations of motion are given by 
\br
\label{t1b}
\frac{d^2}{dt^2} q_1(t) &=& - e^{q_1(t)-q_2(t)} - \zeta_1 \sinh{(\zeta_1 Q_3(t))} - \zeta_2 \sinh{(\zeta_2 Q_3(t))}- \zeta_3 \sinh{(\zeta_3 Q_3(t))},\\
\label{t2b}
\frac{d^2 }{dt^2} q_2(t) &=&  e^{q_1(t)-q_2(t)} -  e^{q_2(t)-q_3(t)}  + 2\zeta_1 \sinh{(\zeta_1 Q_3(t))} + 2 \zeta_2 \sinh{(\zeta_2 Q_3(t))}+ 2\zeta_3 \sinh{(\zeta_3 Q_3(t))},\\
\frac{d^2}{dt^2} q_3(t)&=& e^{q_2(t)-q_3(t)}  - \zeta_1 \sinh{(\zeta_1 Q_3(t))} - \zeta_2 \sinh{(\zeta_2 Q_3(t))}- \zeta_3 \sinh{(\zeta_3 Q_3(t))}. \label{t3b}\\
Q_3(t) &\equiv& q_1(t) - 2 q_2(t) + q_3(t),
\er
where the potential derivative terms $[\zeta_i \sinh{(\zeta_i Q_3)}]$ represent smooth deformations of the $N=3$ Toda chain. Specifically, the potential terms $U^{(i)}$ depends on deformation parameters $\zeta_i$ such that, in the limit $\zeta \rightarrow 0$, it reduces to the standard $N=3$ Toda chain (\ref{t1})-(\ref{t3}). To carry out the expansion around the $N=3$ Toda chain, we consider solutions that are odd under time reflection.

In order to examine the parity versus dynamics interplay for our model let us consider the representative potential of type (\ref{Ui}) with equal parameters 
\br
\zeta_i = \epsilon,\,\,\,\, i=1,2,3.
\er
So, the Hamiltonian becomes
\br
\label{3bodyf}
H= \sum_{n=1}^3 \frac{p_i^2}{2} + e^{q_1-q_2} + e^{q_2-q_3} + 3 \cosh{[\epsilon (q_1-2 q_2+q_3)]},
\er
where $\epsilon$ is the deformation parameter, and the additive constant, $-3$, coming from the potential components (\ref{Ui}) has been dropped. The equations of motion become
\br
\label{t1b1}
\frac{d^2}{dt^2} q_1(t) &=&  -e^{Q_1(t)} - 3 \epsilon \sinh{(\epsilon Q_3(t))},\\
\label{t2b1}
\frac{d^2 }{dt^2} q_2(t) &=& e^{Q_1(t)} -  e^{Q_2(t)}  + 6 \epsilon \sinh{(\epsilon Q_3(t))},\\
\frac{d^2}{dt^2} q_3(t)&=& e^{Q_2(t)}  - 3 \epsilon \sinh{(\epsilon Q_3(t))}, \label{t3b1}\\
Q_3(t) &\equiv& q_1(t) - 2 q_2(t) + q_3(t),\,\,\,Q_1(t) \equiv q_1(t) - q_2(t),\,\,\,Q_2(t) \equiv q_2(t) - q_3(t),\,\,\,
\er
In this case the quasi-conservation law (\ref{ano2})-(\ref{quasi2}) becomes
\br
\label{ano211}
\frac{1}{3} \dot{I}_3 &\equiv& {\cal A}_3(t)\\
 &=& - 3 \epsilon [p_1^2 -2 p_2^2 + p_3^2 - e^{Q_1} -e^{Q_2}] \sinh{(\epsilon Q_3)}.
 \label{quasi211}
\er 
Note that for $\epsilon =0$ the anomaly in (\ref{ano211})-(\ref{quasi211}) vanishes, giving rise to the exact conservation of the charge $I_3$ of the usual Toda chain. Let us consider a small parameter $|\epsilon| <1$. Then, we can expand the coordinates, as solution of the modified equations (\ref{t1b1})-(\ref{t3b1}), in power series in $\epsilon$
\br
\label{pertq}
q_i &=& q_i^0 + \epsilon q_i^1 + \epsilon^2 q_i^2+...; \,\,\,\,\, i=1,2,3.\\
\label{pertQ}
Q_a &=& Q_a^0 + \epsilon Q_a^1 + \epsilon^2 Q_a^2+...; a=1,2.\\
Q_3 &=& Q_3^0 + \epsilon Q_3^1 + \epsilon^2 Q_3^2+...,\\
\label{pertQ12}
Q_1^k &=& q_1^k - q_2^k,\,\,\,\,\,\,\,Q_2^k = q_2^k - q_3^k,\,\,\,\, k=0,1,2,3, ...\\
 Q_3^k&\equiv & q_1^k - 2 q_2^k + q_3^k,\,\,\,\,\,\, k=0,1,2,... 
\er
Using the exact solutions (\ref{sol1}) one can show that
\br
\label{QQ0}
Q_3^0 &=& q_1^0 - 2 q_2^0 + q_3^0,\\
&=& 0.\label{QQ01}
\er
Then, the first four ($k=0,1,2,3$) components of the relevant coordinates $q_i^k$ satisfy
\br
\label{eq0}
\frac{d^2}{dt^2} q_1^0 &=& - e^{Q_1^0},\,\,\,\,\,\,\,\,\,\,\,\,\,\,\,\,\,\,\,\,\,\,\,\,\,\,\,\,\,\,\frac{d^2}{dt^2} q_3^0 =  e^{Q_2^0},\,\,\,\,\,\,\, \,\,\,\,\,\,\,\,\,\,\,\,\,\,\,\,\,\,\,\,\,\,\,\,\,\,\,\,\frac{d^2}{dt^2} q_2^0 =  e^{Q_1^0} -  e^{Q_2^0}\\
\label{eq1}
\frac{d^2}{dt^2} q_1^1 &=& - e^{Q_1^0}\, Q_1^1,\,\,\,\,\,\,\,\,\,\,\,\,\,\,\,\,\,\,\,\,\,\,\frac{d^2}{dt^2} q_3^1 = e^{Q_2^0} \, Q_2^1, \,\,\,\,\,\,\,\,\,\,\,\,\,\,\,\,\,\,\,\,\,\,\,\,\,\,\,  
\frac{d^2}{dt^2} q_2^1 =  e^{Q_1^0}\, Q_1^1 -  e^{Q_2^0}\, Q_2^1,\\
\label{eq2}
\frac{d^2}{dt^2} q_1^2 &=& - e^{Q_1^0} \, (\frac{1}{2} (Q_1^1)^2 + Q_1^2),\\
\frac{d^2}{dt^2} q_3^2 &=& - e^{Q_2^0}\, ( \frac{1}{2} (Q_2^1)^2 + Q_2^2),\\
\frac{d^2}{dt^2} q_2^2 &=& e^{Q_1^0}\,( \frac{1}{2}(Q_1^1)^2 + Q_1^2)-e^{Q_2^0}\,( \frac{1}{2} (Q_2^1)^2 + Q_2^2),\\
\label{eq31}
\frac{d^2}{dt^2} q_1^3 &=& - e^{Q_1^0} \, (\frac{1}{6} (Q_1^1)^3 + Q_1^3+ Q_1^1 Q_1^2) - 3 Q_3^1,\\
\label{eq32}
\frac{d^2}{dt^2} q_3^3 &=& - e^{Q_2^0}\, ( \frac{1}{6} (Q_2^1)^3 + Q_2^3+ Q_2^1 Q_2^2)  - 3 Q_3^1,\\
\label{eq33}
\frac{d^2}{dt^2} q_2^3 &=& e^{Q_1^0}\,(\frac{1}{6} (Q_1^1)^3 + Q_1^3+ Q_1^1 Q_1^2 )-e^{Q_2^0}\,( \frac{1}{6} (Q_2^1)^3 + Q_2^3+ Q_2^1 Q_2^2) + 6 Q_3^1.
\er
Notice that the zero'th order equations (\ref{eq0}) correspond to the usual Toda chain equations (\ref{t1})-(\ref{t3}). The system of equations (\ref{eq1}) for the first order coordinates $q^1_1, q_2^1$ and $q_3^1$ depends on the zero'th order solution appearing as $e^{Q^0_a}\, (a=1,2)$. Similarly, going to the next order equations (\ref{eq2})-(\ref{eq33}) one needs the previous order solutions $Q^k_a\, (a=1,2,\, k=0,1,2)$. 

Notice that the composed coordinate $Q_3$ appears explicitly only at the third order in $\epsilon$ expansion of the equations of motion. In fact, the first appearances of the component $Q_3^1$ ( $Q_3 = Q_3^0+ \epsilon Q_3^1+...$) occur at the last terms of the Eqs. (\ref{eq31})-(\ref{eq33}) corresponding to the third order equations of motion. The component $Q_3^0$ has been neglected due to the result in (\ref{QQ0})-(\ref{QQ01}). However, the effective role of $Q_3^1$ will be further examined below in connection with the solutions for the first-order coordinates $q_i^1$.
 
Since the zero'th order $ Q_3^0(t)$ vanishes identically (\ref{QQ0})-(\ref{QQ01}) when evaluated on the zero'th order $q_i^{0}\,'s$ solution (\ref{sol1}), one can search for the first non-trivial equation of motion for $Q_3^1(t) = Q_1^1(t)-Q_2^1(t) = q_1^1 - 2 q_2^1 + q_3^1$. So, from the Eqs. (\ref{eq1}) and using the zero'th order $q_i^{0}\,'s$ solution, one has 
\br
\label{QQ1}
\frac{d^2}{dt^2}  Q_3^1(t) + 3 \,\mbox{sech}[\frac{t}{\sqrt{2}}]^2 Q_3^1(t) = 0.
\er
The Eq. (\ref{QQ1}) is a classical P\"oschl–Teller type ordinary differential equation. It has the next two independent solutions \cite{abramowitz, flugge}
\br
 Q^1_{3, A}(t) &=& \frac{1}{2} [3\, \tanh^2{[\frac{t}{\sqrt{2}}]} -1],\\
\label{Q1B}
  Q^1_{3, B}(t) &=& \frac{1}{4} [3\, \tanh^2{[\frac{t}{\sqrt{2}}]} -1] \log{\frac{1+ \tanh{[\frac{t}{\sqrt{2}}]} }{1-\tanh{[\frac{t}{\sqrt{2}}]}}} - \frac{3}{2} \tanh{[\frac{t}{\sqrt{2}}]}.
\er
Note that, only the second solution $Q^1_{3, B}(t)$ satisfies the odd parity condition $Q^1_{3, B}(-t)=-Q^1_{3, B}(t)$. 

Since $Q_1^0 =Q_2^0 = \mbox{sech}^2(\frac{t}{\sqrt{2}})$ and $Q_3^1(t) = Q_1^1(t)-Q_2^1(t)$, the first order equation for $q_2^1(t)$ from (\ref{eq1}) can be written as   
\br
\label{eq12}
\frac{d^2}{dt^2} q_2^1(t) = \mbox{sech}^2(\frac{t}{\sqrt{2}})  Q^1_{3, B}(t),
\er
where we have inserted $Q^1_{3, B}(t)$ given in (\ref{Q1B}), which is the odd solution of (\ref{QQ1}).
Next, the solution of (\ref{eq12}) satisfying $q_2^1(-t) = - q_2^1(t)$ becomes
\br
q_2^1 &=& \frac{1}{4} \mbox{sech}^2(\frac{t}{\sqrt{2}}) [\sqrt{2} \, t + \sinh(\sqrt{2} \,t)].\label{q21sol}
\er 
The both coordinates $q_a^1\, (a=1,3)$ satisfy the next differential equation
\br
\label{q121}
\frac{d^2}{dt^2} q_a^1(t) + \mbox{sech}^2[\frac{t}{\sqrt{2}}] \, q_a^1(t) - \frac{1}{4} \mbox{sech}^4(\frac{t}{\sqrt{2}}) [\sqrt{2} \, t + \sinh(\sqrt{2} \,t)] &=&0,\,\,\,\,\, a= 1,3. 
\er
The general solution of this second order non-homogeneous differential equation has the general solution
\br
q_a^1(t) = C_1 \tanh{(\frac{t}{\sqrt{2}})} + C_2 [\frac{t}{\sqrt{2}} \tanh{(\frac{t}{\sqrt{2}})} -1] + \frac{1}{8} \mbox{sech}^2(\frac{t}{\sqrt{2}}) [\sqrt{2} \, t \cosh{(\sqrt{2} t)} - \sinh{(\sqrt{2} t)}], \,\,\,\, a=1,3,
\label{q13sol}
\er
with $C_1, C_2$ being constants. However, for our purposes below, we choose the solution with $C_2=0$, which furnishes odd parity solutions $q_a^1(-t) = -q_a^1(t)$ for $a=1,3$.

The Fig. 2 shows the plots of  $q_1(t) = q_1^0+ \epsilon q_1^1(t), q_2(t) = q_2^0+ \epsilon q_2^1(t)$ and $q_3(t) = q_3^0+ \epsilon q_3^1(t)$, for $\epsilon=0.5$. Indeed, Fig. 2 presents the coordinates corresponding to the modified model, providing a first-order correction in $\epsilon$ to those shown in Fig. 1, such that the symmetry condition (\ref{sym1}) is preserved under the deformation.

\begin{figure}
\centering
\label{fig2}
\includegraphics[width=1.5cm,scale=4, angle=0,height=6cm]{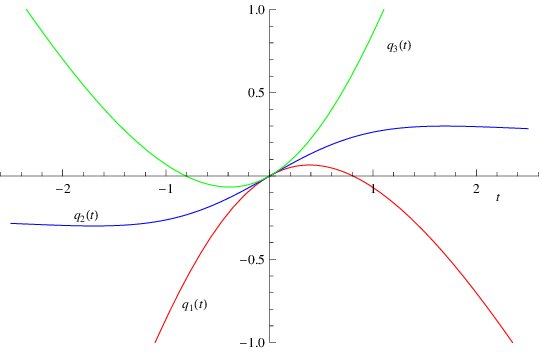} 
\parbox{6in}{\caption{(color online) The $N=3$ modified Toda chain coordinates $q_1(t)$ (red), $q_2(t)$ (blue) and $q_3(t)$ (green) up to the first order contribution for $\epsilon=0.5$. Notice that the symmetry (\ref{sym1}) is realized.}}
\end{figure}

Next, let us expand the anomaly ${\cal A}_3(t)$ in (\ref{ano211})-(\ref{quasi211}) in powers of $\epsilon$
\br
\nonumber
{\cal A}_3(t) &=& -3 \epsilon^2 \Big[(p_1^0)^2 -2 (p_2^0)^2 + (p_3^0)^2 - e^{Q^0_1} -e^{Q^0_2} + \\
&& \epsilon [2(p_1^0 p_1^1 -2 p_2^0 p_1^1+ p_3^0 p_3^1) - e^{Q^0_1} Q_1^1 -e^{Q^0_2} Q_2^1] \Big]  \(Q_3^0 + \epsilon Q_3^1 + \epsilon^2 Q_3^2 \), \label{eps3}\\
 &=& -3 \epsilon^3 [(p_1^0)^2 -2 (p_2^0)^2 + (p_3^0)^2 - e^{Q^0_1} -e^{Q^0_2}] Q_3^1 - \nonumber\\
 && 3 \epsilon^4 \Big[ [2(p_1^0 p_1^1 -2 p_2^0 p_1^1+ p_3^0 p_3^1) - e^{Q^0_1} Q_1^1 -e^{Q^0_2} Q_2^1] Q_3^1 + \nonumber\\
  &&  [(p_1^0)^2 -2 (p_2^0)^2 + (p_3^0)^2 - e^{Q^0_1} -e^{Q^0_2}] Q_3^2    \Big]+
{\cal O}(\epsilon^5),\,\,\,\,\,\,\,\,Q_3^1 = q_1^1 - 2 q_2^1 + q_3^1, \label{eps31}
\er
where we have used $Q_3^0=0$ given in (\ref{QQ0})-(\ref{QQ01}). Note that the lowest order correction to the anomaly becomes of the order $\epsilon^3$. This observation is consistent with the previous discussion indicating that the first-order composite coordinate 
$Q_3^1$ starts to contribute to the dynamics only at order 
$\epsilon^3$ (see the last terms of (\ref{eq31})-(\ref{eq33})), since the zeroth-order term $Q_3^0$ vanishes when evaluated on the Toda chain solution. Nevertheless, one may note that this composite coordinate effectively appears at first order in  $\epsilon$ in Eq. (\ref{eq1}). Its role becomes evident upon rewriting the coordinates as $Q_1^1 = Q_3^1 +Q_2^1$ and $Q_2^1=Q_1^1 - Q_3^1$, revealing the interplay between the first-order components and the composite coordinate $Q_3^1$. This statement is further supported by the emergence of the differential equation (\ref{QQ1}) for the composite quantity $Q_3^1$, whose solutions play a crucial role in constructing the coordinate components
 $q_i^1$ in (\ref{q21sol}) and (\ref{q13sol}).   

Notice that the above perturbative expansion of the anomaly relies on the result $Q_3^0=0$ given in (\ref{QQ0})-(\ref{QQ01}), which is related to the CM coordinate system at rest. However, in the laboratory coordinate system one can get a nonvanishing zero'th order contribution to the composite coordinate $Q_3^0$. In fact, in this case from (\ref{Q3y}) one must have 
\br
\label{Q30cm1}
Q_3(t) &=& 3 Q_{CM}(t) - 3 q_2(t)\\
	&=& 3 (v_o \,t + Q_{o} -  q_2(t)), \label{Q30cm}
\er 
where $v_o$ is the CM velocity and $Q_{o}$ is an integration constant of the second order equation (\ref{CM2}). Therefore, one can argue that, at the zero'th order solution, relation (\ref{Q30cm}) implies that the leading correction to the anomaly in (\ref{eps3}) is of order $\epsilon^2$. Nevertheless, this contribution still exhibits the odd-parity symmetry under time refection, provided that $Q_{o}=0$. This shows that for generic Toda trajectories with $Q^0_3(t) \neq 0$ the power counting is modified.  
 
Next, substituting the relevant quantities into (\ref{eps31}), and considering the lowest order contribution to the anomaly, one has  
\br
\label{anolast}
{\cal A}_3(t)&=& -6 \epsilon^3 \Big[2-3 \mbox{sech}^2[\frac{t}{\sqrt{2}}]\Big] Q_{3,B}^1,
\er
where we have inserted the odd parity solution $Q_{3,B}^1$.
Taking into account $Q_{3,B}^1$ in (\ref{Q1B}), the time integral of the last expression (\ref{anolast}) becomes
\br
\int_{t_i}^{t_f} {\cal A}_3(t) dt &=& -\frac{3}{2} \epsilon^3 \Big\{ \sqrt{2} (t_f^2-t_i^2) + 3 \sqrt{2}[\mbox{sech}^2(\frac{t_i}{\sqrt{2}})-\mbox{sech}^2(\frac{t_f}{\sqrt{2}})] + \\
&& 3 t_i \tanh{(\frac{t_i}{\sqrt{2}})} [1+ \tanh^2{(\frac{t_i}{\sqrt{2}})}] 
- 3 t_f \tanh{(\frac{t_f}{\sqrt{2}})} [1+ \tanh^2{(\frac{t_f}{\sqrt{2}})}] \Big\}.
\er
Evaluating this expression for large $t_0 > 0$, one has
\br
\label{analy1}
\lim_{\begin{array}{c}
t_i \rightarrow - t_0 \\
t_f \rightarrow t_0\end{array}} \int_{t_i}^{t_f} {\cal A}_3(t) dt =0.
\er
So, one has a vanishing anomaly up to the order  $\epsilon^3$. Below, we will check through numerical simulation of the modified Toda  chain the vanishing anomaly for moderate values of $|\epsilon|<1$.         

One may interpret that the third conserved charge 
$I_3$, initially defined at 
$t=-t_0$, $I_3(-t_0)$, attains the same value after the system evolves dynamically to $t=+t_0$. This occurs under the condition that the coordinates vanish at the symmetric point $t=0$, namely $q_1(t=0)=q_2(t=0)=q_3(t=0)=0$. Throughout this temporal evolution, the configuration preserves time-reflection symmetry, ensuring that the time   behavior of $I_3$ remains invariant under the transformation $t\rightarrow -t$, $I_3(-t_0) = I_3(t_0)$. This behavior can be appreciated qualitatively in the Fig. 3. In this Fig. the energy $E(t)$ and the third charge 
$I_3$ are plotted for $\epsilon =0.12$. They are evaluated for the coordinates $q_i(t) = q_i^0+ \epsilon q_i^1,\, i=1,2,3$. Notice that the charge $I_3(t=\pm t_0)$ is asymptotically constant for large $|t|= t_0 > 2.5$. In fact, it takes the value $\lim_{|t| \rightarrow +\infty}[I_3(t)]= 0.501$. The validity of this result is examined below by means of numerical simulations.

\begin{figure}
\centering
\label{fig3}
\includegraphics[width=1.5cm,scale=4, angle=0,height=8cm]{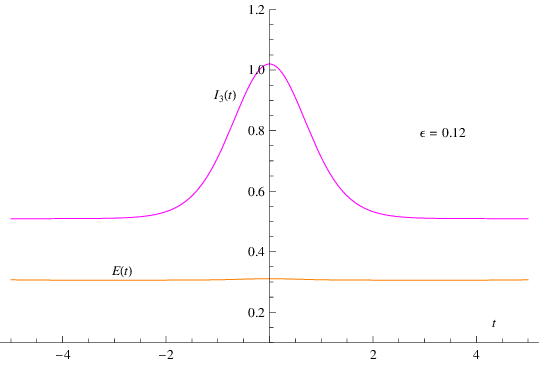} 
\parbox{6in}{\caption{(color online)  Plots of energy $E(t)$  and the third charge $I_3$ for $\epsilon =0.12$ evaluated for the coordinates $q_i(t) = q_i^0+ \epsilon q_i^1,\, i=1,2,3$. Notice the asymptotically constant $I_3(t=\pm t_0) = 0.5$ behavior for large $|t|= t_0 > 2.5$.}}
\end{figure}

\section{Numerical analysis of the modified $N=3$ open Toda model}
\label{sec:num}

To explore the impact of the deformation on the integrable structure of the open Toda chain, we performed numerical simulations of the modified system with Hamiltonian (\ref{3bodyf}), where the last term introduces a nonlinear coupling controlled by the deformation parameter 
$\epsilon$. For $\epsilon =0$, the model reduces to the standard open Toda chain, which is completely integrable. As $\epsilon$ increases, the additional potential term breaks exact integrability but preserves key qualitative features such as the momentum and energy conservation.

The deformation term couples the discrete `curvature' $Q_3 =q_1-2q_2+q_3 $ to the Toda dynamics. To obtain stationary and symmetric configurations, we applied the relaxation method (see Appendix \ref{sec:apprelax}), using the first order perturbative $N=3$ deformed Toda chain profiles $q_i(t) = q^0_i(t) + \epsilon q_i^1(t),\, i=1,2,3$ in (\ref{pertq}) as initial inputs (these input coordinates are plotted in Fig. 2). The time is discretized with the step $dt = 0.001$, and the boundary conditions $q_i(t=0) = 0,\,q_i(t=T) \neq 0$ are imposed at the origin $t=0$ and at $t = T$, respectively, with $T\rightarrow large$. The iterative scheme converged efficiently for small and moderate values of $|\epsilon| < 1$, after less than $K = 10$ iterations. Indeed, the small number of iterations needed for convergence is due to the quality of the initial guess, and the deformation nonlinearity strength being of  hyperbolic type. This  yields a stable, reflection-symmetric solutions satisfying $q_1(-t) = -q_3(t),\,q_2(-t) = -q_2(t)$. The relaxation method is discussed in the appendix \ref{sec:apprelax}.
 
Our numerical results are presented in the Fig. 4. This Fig. shows the numerical solution for $N=3$ modified Toda chain coordinates $q_1(t)$ (red), $q_2(t)$ (blue) and $q_3(t)$ (green) for $\epsilon=0.75$. Note that $q_2(t=\pm T) \approx \pm \frac{\epsilon}{2}$. For small $|\epsilon| <1$, the numerical solutions remain very close to the analytical solutions obtained in the first order in $\epsilon$ perturbative regime (\ref{pertq}), which were plotted in the Fig. 2 for $\epsilon =0.5$. One can notice qualitatively that the symmetry (\ref{sym1}) of the coordinates under time-reflection is realized in the Fig. 4. The dark line shows the anomaly ${\cal A}_3(t)$ (\ref{ano211})-(\ref{quasi211}). Observe  that qualitatively  the odd symmetry (\ref{odd1}) of the anomaly is realized. We performed a numerical  integration $\int_{-t_0}^{t_0}\, dt\,  {\cal A}_3,\, (t_0 = 5)$, as in (\ref{ano011}), which provides a vanishing quantity with numerical accuracy of the order of $10^{-15}$. Then, the vanishing anomaly condition (\ref{vanish}) is verified.

Additionally, the numerical computation of the deformed third invariant $I_3 (t)$ for $\epsilon = 0.12$ yields a profile that closely reproduces that of Fig. 3, with no discernible differences within numerical accuracy. So, numerical simulations reveal that the deformed charge  $I_3(t)$ exhibits noticeable variations around the central region of the time evolution $t \approx 0$. As time increases, these fluctuations diminish, and $I_3(t)$ gradually returns to the value attained in the distant past, indicating that the charge remains effectively conserved in the asymptotic regime.   
   
Overall, the numerical analysis highlights how the interplay between exponential and hyperbolic deformation  governs the emergence of quasi-integrable regimes. This behavior provides a useful framework to study deformed extensions of Toda-type systems, where similar balance between nonlinearity and deformation-induced coupling may yield rich dynamical phenomena.

\begin{figure}
\centering
\label{fig4}
\includegraphics[width=1.5cm,scale=4, angle=0,height=8cm]{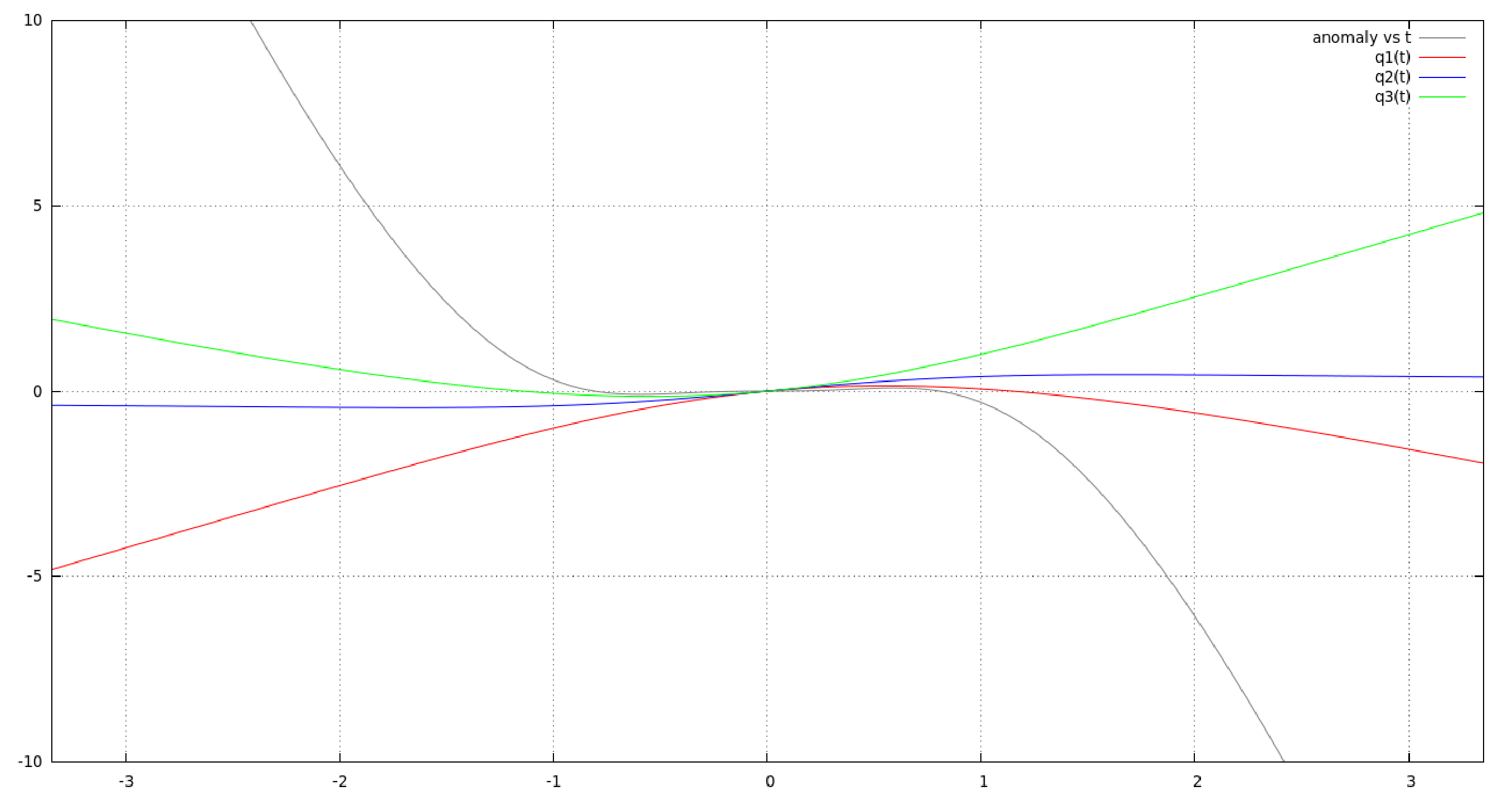} 
\parbox{6in}{\caption{(color online) The numerical solution for $N=3$ modified Toda chain coordinates $q_1(t)$ (red), $q_2(t)$ (blue) and $q_3(t)$ (green) for $\epsilon=0.75$. Notice that the symmetry (\ref{sym1}) of the coordinates is realized. The dark line shows the anomaly ${\cal A}_3(t)$ with the odd symmetry (\ref{odd1}).}}
\end{figure}

\section{Conclusions and discussions}
\label{sec:conclu}

We have analyzed a quasi-integrable deformation of the three-particle open Toda chain induced by a translation-invariant three-body interaction. While this deformation breaks the exact Toda chain  integrability, it preserves energy and momentum and leads to an exact quasi-conservation law for the cubic invariant 
$I_3$. It has been verified, analytically (\ref{analy1}) and numerically (section \ref{sec:num}), that the time integrated anomaly $\int_{-t_0}^{t_0}\, dt  {\cal A}_3$ in Eqs. (\ref{ano211})-(\ref{quasi211}) vanishes under specific time-reflection symmetry conditions (\ref{sym1})-(\ref{potsymm1}). This establishes  a non-perturbative realization of quasi-integrability, since this result implies (\ref{i30}), i.e.  $ I_3(t_0) = I_3(-t_0)$. So, in the limit $t_0 \rightarrow \infty$, $I_3$ becomes an asymptotically conserved charge which characterizes the $N=3$ deformed Toda chain as a quasi-integrable model. The behavior of $I_3(t)$ for large time $t$ can be observed in the Fig. 3.

So, this framework provides a minimal setting to study how integrable dynamics deform into quasi-integrability, offering insight into the persistence of asymptotically conserved structures beyond the fully integrable regime. 

In section \ref{sec:num}, we have also carried out numerical simulations for a specific form of the deformation potential $U_3$ which defines the deformed $N=3$ open Toda chain Hamiltonian (\ref{3bodyf}), focusing particularly on the long-time dynamics associated with the asymptotically conserved charge $I_3$. The numerical solution for $N=3$ modified Toda chain coordinates $q_1(t)$ (red), $q_2(t)$ (blue) and $q_3(t)$ (green) for $\epsilon=0.75$ have been plotted in the Fig 4. Notice in this Fig. that the time-reflection symmetries (\ref{sym1}) and (\ref{odd1}), respectively, of the coordinates and the anomaly ${\cal A}_3(t)$, were realized numerically.

The perturbative analytical prediction of a vanishing anomaly was corroborated by numerical results obtained via the relaxation method, in which the analytical first-order solution in the deformation parameter $\epsilon$ was employed as the initial trial input.

The lowest-order correction to the anomaly appears at order 
$\epsilon^3$ and it is proportional to the coordinate $Q_3^1$ (see \ref{eps31}). However, it relies on expanding around a Toda trajectory with $Q^0_3(t) = 0$. For a generic Toda solution with $Q_3(t)^0 \neq  0$ as in (\ref{Q30cm}), the lowest-order correction to the anomaly starts at $\epsilon^2$. The composite coordinate $Q_3$ is the argument of the deformation potential, $3 \cosh{(\epsilon Q_3)}$, in (\ref{3bodyf}). Although $Q_3^1$ formally arises at the order 
$\epsilon^3$ in the Eqs. of motion (\ref{eq31})-(\ref{eq33}), its dynamical role becomes evident already at the first order level. This behavior is further supported by the differential equation (\ref{QQ1}) for $Q_3^1$, whose solution determine the coordinate components $q_i^1$ in Eqs. (\ref{q21sol}) and (\ref{q13sol}). 

Finally, generalizations of the present analysis to the open and closed Toda chain with arbitrary number $N$ of particles constitute a natural direction for future investigation. 

\[\]
\noindent {\bf Acknowledgements}

We thank M. C. de Oliveira, N. A. de Almeida and E. S. Marinho for useful discussions.  
  
\[\]

{\bf Funding information}: This study has not received any financial support.
 
\appendix

\section{Relaxation method}

\label{sec:apprelax}

We perform the numerical simulation of the second order nonlinear system of equations (\ref{t1b1})-(\ref{t3b1}) using the relaxation method. Notice that this system of equations is invariant under the symmetry transformations (\ref{tcr}). So, in order to find the solutions which satisfy the symmetry (\ref{tcr}) it is enough to solve the system of equations in the interval $ t \in [0, T]$ for $T$ being a large time. Therefore, in the framework of the relaxation method we must impose six conditions at the two boundaries $t = 0$ and $t = T$. At $t = 0$ we impose three conditions
\br
\label{bc0}
q_i(t=0)=0,\,\,\,\, i=1,2,3.
\er
And at $t = T$ we choose the next three conditions
 \br
q_2(T) &=& q_{20}=const.,\label{bc1}\\
\label{bc2}
q_1(T) &=& 3\epsilon v_o T + (1-3 \epsilon) q_{20} -\log{(6|\epsilon|)},\\
\label{bc3}
q_3(T) &=& 3(1-\epsilon) v_o T + (3 \epsilon-2) q_{20} +\log{(6|\epsilon|)},\\ 
Q_{CM}(T) &=& \frac{1}{3} (q_1(T) + q_2(T) + q_3(T)),\nonumber
\\ &=& v_o T.\label{bc123}
\er 
Observe that the these conditions fix the values of the corresponding coordinates at a specified late time $t=T$, with $T \rightarrow large$. Moreover, these conditions describe asymptotic states of free particles undergoing uniform motion along rectilinear trajectories. The b.c. for $q_2(t)$ is motivated by the observation that its first order in $\epsilon$ solution tends asymptotically to a constant, i.e. $q_2(t \rightarrow \pm T) \rightarrow \pm \frac{\epsilon}{2}$ (see Fig.2). Whereas, in the usual Toda lattice it is a trivial solution $q_2(t)=0$ (\ref{sol1}) (see Fig. 1). 

The boundary conditions (\ref{bc1})-(\ref{bc3}) are consistent with the asymptotic limits of the equations of motion (\ref{t1b1})-(\ref{t3b1}) at $t\rightarrow T$, with $T$ large. In particular, they satisfy the equation for the CM coordinate $Q_{CM}(t)$ in (\ref{CM1})-(\ref{CM2}) evaluated at $t=T$ and $Q_o=0$, as in (\ref{bc123}).  
      
The relaxation method is a well-established numerical procedure for handling boundary conditions of this type. As shown below, the method is initiated with an appropriately selected trial (initial guess) function and proceeds by iteratively constructing successive approximations to the solution. After a sufficient number of iterations, denoted by $K$, the resulting sequence is found to converge to the exact (true) solution.

We will check the accuracy of our numerical method by comparing to the exact analytical results obtained in sec. \ref{sec:hirota}. In this case one considers the b.c.'s: $q_1(T) = - q_3(T) = \log{4} - \sqrt{2}\, T,\,\,\, q_2(T) =0$ for $T$ large, and $q_i(t=0) = 0\, (i=1,2,3)$, suitable for the undeformed Toda lattice ($\epsilon =0$). These b.c.'s  are consistent with the analytical Hirota-Moser solutions (\ref{sol1}).   

We now express the system (\ref{t1b1})-(\ref{t3b1}) collectively as a set of second-order nonlinear differential equations
\br
\label{eq11r}
\frac{d^2}{dt^2} q_i(t) - W_i(q_1, q_2, q_3; \epsilon)=0,\,\,\, i=1,2,3,
\er
where the $W_i(q_1, q_2, q_3; \epsilon)'s$ denote the r.h.s.'s of the system of equations (\ref{t1b1})-(\ref{t3b1}), respectively.
  
Next, let us  define the following  discretizations
\br
t_n &=& t_1 + (n-1) dt,\,\,\,\,\,\,dt = \frac{T}{M-1},\,\,\, n =1,2,3,....M,   \\
q_{i,n} &=&  q_i(t_n),\,\,\,\, t_1 =0,\,\,\,\, t_M = T,
\er
where $M$ ($M >> 2$) is an integer and $dt$ defines the grade spacing $dt = t_{n+1}-t_n$.  We use the central difference approximation of the second order derivative  
\br
\label{diff}
\frac{d^2}{dt^2} q_i(t) \approx   \frac{q_{i, n+1}- 2 q_{i,n} + q_{i,n-1}}{2 \,dt^2}.
\er
Next, let us approximate  the  differential eqs.  (\ref{eq11r}) through the finite difference method. So, making use of the approximation  (\ref{diff}) one has 
\br
\frac{q_{i, n+1}- 2 q_{i,n} + q_{i,n-1}}{2 \,dt^2} - W_i(q_{1,n}, q_{2,n}, q_{3,n}; \epsilon)= 0,\,\,\,\,\, i=1,2,3.
\er 
Therefore, one can write  
\br
\label{iter}
q_{i,n} &=& \Big[\frac{q_{i, n+1} + q_{i,n-1}}{2} - dt^2\, W_i(q_{1,n}, q_{2,n}, q_{3,n}; \epsilon)\Big],\,\,\,\,\, n=2,3...,M-1.\\
q_{i,1} &=&0,\,\,\,\,q_{i,M} = q_i(T),\,\,\,\,\,i=1,2,3,\label{iter1}
\er
where in the last line we have imposed the b.c.'s (\ref{bc0}) and (\ref{bc1})-(\ref{bc3}).
 
Some comments are in order here.  

First, Eqs. (\ref{iter})-(\ref{iter1}) will be employed to carry out the iterative construction of the solutions for the arrays 
$q_{i,n}$, as is standard in the relaxation method.   

Second, one starts by assuming an initial guess function for $q_{i,n} \equiv q_{i,n}^{(0)} $ and assigning  those values on the right hand side of (\ref{iter}) to the variables $q_{i,n}$ in order to construct the first iterated array $q_{i,n} \equiv  q_{i,n}^{(1)}$ on the left hand side. An initial trial solution $q_{i,n}^{(0)}$ is taken from the set of analytic  solutions (\ref{pertq}) of order $\epsilon^1$.  
  
Third, the process above must be repeated $K$ times until the set of solutions $q_{i,n}^{(K)}$ converge to the final solution $q_{i,n}$, within numerical accuracy.

\end{document}